\documentclass[11pt]{article}

\usepackage{jheppub}

\usepackage{amsmath,amssymb,amsthm,cancel,hyperref,graphicx,xcolor}
\usepackage{mathrsfs,wasysym}
\usepackage{booktabs}
\usepackage{tikz}
\usetikzlibrary{calc}
\usepackage{placeins}
\usepackage{enumitem}

\usepackage{amsmath}

\usepackage{slashed}
\usepackage{graphicx}

\usepackage{soul}

\DeclareRobustCommand{\Sec}[1]{Sec.~\ref{#1}}

\DeclareRobustCommand{\Fig}[1]{Fig.~\ref{#1}}

\DeclareRobustCommand{\Eq}[1]{Eq.~(\ref{#1})}

\DeclareRobustCommand{\InRef}[1]{Ref.~\cite{#1}}
\DeclareRobustCommand{\Refs}[1]{Refs.~\cite{#1}}

\newcommand{\be}{\begin{eqnarray}}
\newcommand{\ee}{\end{eqnarray}}
 
\allowdisplaybreaks

\usepackage{color}
\definecolor{darkblue}{rgb}{0,0,0.5}
\definecolor{darkgreen}{rgb}{0,0.5,0}

\title{A Step Toward Interpretability: Smearing the Likelihood}
\author{Andrew J.~Larkoski}
\affiliation{American Physical Society, Hauppauge, New York 11788, USA}

\emailAdd{larkoski@aps.org}

\abstract{
The problem of interpretability of machine learning architecture in particle physics has no agreed-upon definition, much less any proposed solution.  We present a first modest step toward these goals by proposing a definition and corresponding practical method for isolation and identification of relevant physical energy scales exploited by the machine.  This is accomplished by smearing or averaging over all input events that lie within a prescribed metric energy distance of one another and correspondingly renders any quantity measured on a finite, discrete dataset continuous over the dataspace.  Within this approach, we are able to explicitly demonstrate that (approximate) scaling laws are a consequence of extreme value theory applied to analysis of the distribution of the irreducible minimal distance over which a machine must extrapolate given a finite dataset.  As an example, we study quark versus gluon jet identification, construct the smeared likelihood, and show that discrimination power steadily increases as resolution decreases, indicating that the true likelihood for the problem is sensitive to emissions at all scales.
}

\begin{document}
\maketitle

\section{Introduction of Interpretability of Machine Learning}

While machine learning has become a dominant tool of particle physics, see \Refs{Larkoski:2017jix,Kogler:2018hem,Guest:2018yhq,Albertsson:2018maf,Radovic:2018dip,Carleo:2019ptp,Bourilkov:2019yoi,Schwartz:2021ftp,Karagiorgi:2021ngt,Boehnlein:2021eym,Shanahan:2022ifi,Plehn:2022ftl,Nachman:2022emq,DeZoort:2023vrm,Zhou:2023pti,Belis:2023mqs,Mondal:2024nsa,Feickert:2021ajf,Larkoski:2024uoc,Halverson:2024hax} for an incomplete list of recent reviews, there is a dark side to this revolution.  In many situations where machine learning has supplanted ``traditional'' analyses such as observable construction for binary discrimination problems, an understanding of the physics that is exploited, whether at a human intuitive level or from first-principles theoretical calculations, is lacking or even completely absent.  More generally, there is a program in machine learning of interpretability \cite{rudin2019stop,molnar2020interpretable} of how and what a machine learns, where information is stored within its architecture, and how it responds to stimuli or data outside of its training set.  Particle physics would seem to be in a privileged position amongst almost all other realms in which machine learning is used, and perhaps because of underlying theory, nearly-perfect simulated data, enormous (and growing) experimental datasets, one might expect that interpretation would come easy.  However, it is far from true, and even in particle physics there is not an agreed-upon definition of what ``interpretability'' would look like or what one would want it to be; see, e.g., \Refs{Grojean:2022mef,jesseint,mattint} for three recent reviews and talks about this.

Nevertheless, some ideas of interpretability explored within the computer science literature have recently been employed to study machine learning as applied to particle physics tasks.  One example is Shapley values \cite{shapley1951notes,roth1988shapley}, which quantify the amount of credit that each individual amongst an ensemble should be given for accomplishing a particular goal.  Within the context of machine learning for particle physics, Shapley values and related techniques have been applied to determine the observable or observables that provide the greatest separation for a binary discrimination problem, e.g., \Refs{Chang:2017kvc,Roxlo:2018adx,Faucett:2020vbu,Grojean:2020ech,Das:2022cjl,Bhattacherjee:2022gjq,Munoz:2023csn,Chowdhury:2023jof,Englert:2024ufr,Bose:2024pwc}.  While these approaches quantify the importance of one observable over another to discrimination, they do depend on the initial ensemble of observables that one considers.  If an ensemble of observables is not sufficiently expressive for the task at hand, it may be that the optimal or most powerful observable cannot be represented and therefore is completely missed by this approach.  We would therefore want our interpretation framework to be independent of any observable set or ensemble, and be able to identify the truly optimal observable, regardless of the representation of the data we choose.

In this paper, we provide a first step toward interpretability within the space of machine learning as it is used in particle physics.  For concreteness, we will focus on the problem of binary discrimination, but it will be clear that this approach simply generalizes widely.  As such, the goal of a machine learning architecture for binary discrimination of signal $s$ from background $b$ events is to estimate the likelihood ratio, which by the Neyman-Pearson lemma \cite{Neyman:1933wgr} is the observable whose contours maximize signal, for a fixed background contamination.  As a function of the phase space coordinates $\vec x$ on which data lives, the likelihood ${\cal L}$ is simply the ratio of background to signal distributions:
\begin{align}
{\cal L}(\vec x) = \frac{p_b(\vec  x)}{p_s(\vec x)}\,.
\end{align}
The challenge of machine learning lies in the facts that typically the dimensionality of $\vec x$ is large (in particle physics, typically of order of hundreds) and that the data on which the machine is trained are discrete and finite; i.e., some set $\{\vec x_i\}_{i=1}^n$, for $n$ events.  As such, direct evaluation of the likelihood on the training data is not possible, and instead a machine learns a continuous functional form for the likelihood that is designed to extrapolate between points in training data as robustly as possible.\footnote{The statement ``extrapolate between points'' may sound odd, and like a machine would actually be doing {\it interpolation}.  Additionally, machines interpolate rather well, and extrapolate rather poorly, in general, so it may seem like we are selling the strengths of the machine short.  However, interpolation versus extrapolation can change depending on the representation of the data and how one works to fit it.  For example, in real space it may seem like a machine is interpolating, but the machine works to fit the data in a conjugate function or ``momentum space'', in which long-distance correlations are well-described by slowly-varying functions.  By contrast, short-distance correlations are described by high momentum or rapidly-varying functions, and the minimal distance between data points sets an upper bound on the highest possible momentum that can be meaningfully represented.}

The art of machine learning is the way in which this extrapolation is done, the way assumptions of structures in the training data are used, or the assumed functional form and parameters that the machine learns.  However, given a trained machine for some discrimination problem, as a human simply staring at the learned function is almost certainly not useful or enlightening, because modern architectures have millions (or more) parameters and use enormous linear combinations of compositions of compositions of functions in their fitting.  In its raw ``machine'' form, such a function would definitely not be human interpretable, but what that should mean or how simple is simple enough for a human has no obvious definition.

Instead, we will take an approach from the completely opposite direction.  We will advocate for extending the likelihood by a single parameter, a resolution energy scale $\epsilon$, and by varying $\epsilon$, one can study how physics at different scales affects the likelihood.  Specifically, we advocate for the following definition of interpretability; or, that the following should be a part of any broader definition of interpretability in particle physics. We define:
\begin{quote}
{\bf Definition:} ``Interpretability of machine learning in particle physics'' means the isolation and identification of the relevant physical energy scales learned and exploited by the machine.
\end{quote}
In a physics language, we might call this a Wilsonian approach \cite{Wilson:1983xri} to interpretability by which we smear or integrate over our ignorance of physics at short dataspace distances and study the consequences of that smearing on long dataspace distance physics.  We re-emphasize that this definition of interpretability is defined exclusively in terms of physics content, and makes no reference to any possible machine learning architecture, nor its internal weights, nor its internal decision logic.

To do this requires a metric $d(\cdot,\cdot)$ on the dataspace, as a function of phase space points $\vec x$.  $d(\cdot,\cdot)$ then necessarily satisfies the requirements of a metric:
\begin{enumerate}
\item $d(\vec x,\vec x' )\geq 0$\,,
\item $d(\vec x,\vec x' )= 0$ iff $\vec x = \vec x'$\,,
\item $d(\vec x,\vec x' )=d(\vec x',\vec x)$\,,
\item $d(\vec x,\vec x' )+d(\vec x',\vec x'' )\geq d(\vec x,\vec x'' )$\,,
\end{enumerate}
for three points on phase space, $\vec x,\vec x', \vec x''$.  The fourth property is of course the triangle inequality.  Further, for maximal interpretability as a physical energy scale distance, we also enforce the physically-motivated properties:
\begin{enumerate}[resume]
\item $[d(\vec x,\vec x')] = [\text{energy}]$. That is, the units of the metric is energy.
\item $d(\cdot,\cdot)$ is infrared and collinear (IRC) safe \cite{Kinoshita:1962ur,Lee:1964is,Ellis:1996mzs}. That is, $d(\vec x,\vec x') \to 0$ if $\vec x$ and $\vec x'$ differ only by exactly collinear or exactly 0 energy emissions.
\end{enumerate}
A large number of IRC safe particle physics event space metrics have been proposed in recent years, see, e.g., \Refs{Komiske:2019fks,Mullin:2019mmh,CrispimRomao:2020ejk,Cai:2020vzx,Larkoski:2020thc,Tsan:2021brw,Cai:2021hnn,Kitouni:2022qyr,Alipour-Fard:2023prj,Larkoski:2023qnv,Davis:2023lxq,Ba:2023hix,Craig:2024rlv,Cai:2024xnt,Gambhir:2024ndc}, and so there are in principle other desired properties that one can enforce to further select.  In this paper, we will be extremely pragmatic, and use the $p=2$ Spectral Energy Mover's Distance \cite{Larkoski:2023qnv,Gambhir:2024ndc} for the reasons that it is invariant to event isometries, expressible in closed-form, and evaluates extremely fast.  We will review the properties of the Spectral Energy Mover's Distance in \Sec{sec:semd}, but for now will just use the general notation $d(\cdot,\cdot)$ for the metric.

Additionally, we want to emphasize the importance and necessity of these physical requirements on the metric.  The property of IRC safety of the metric is not simply an issue of practical concern, so that predictions of pairwise event distances can be calculated within the perturbation theory of quantum chromodynamics, for example.  IRC safety ensures that events that differ by radiation that in no way modifies the measurable energy flow of a jet are indiscernible by the metric.  This is centrally crucial for our approach to interpretability, and is precisely the property that guarantees that the metric distance is interpretable as we propose.  If the metric were not IRC safe, then, at the very least, experimentally indiscernible events would not necessarily be indiscernible with the metric, and no robust physical conclusions could be made. 

Now, given a metric on the space of events, we can then define a smeared or averaged distribution in which all events within an energy distance $\epsilon$ from a phase space point $\vec x$ of interest are summed.\footnote{Previous methods for smearing over the high-dimensions of the full jet phase space to a human-interpretable low dimensional phase space had been introduced, e.g., \Refs{Datta:2017rhs,Larkoski:2019nwj,Kasieczka:2020nyd}.  However, in many ways these are suboptimal from theoretical and machine learning perspectives because they explicitly project the jet onto a phase space of fixed dimensionality.  Similarly, standard data analysis techniques like $k$-means or $k$-medioids are suboptimal because there is no natural $k$ to choose on data that is approximately scale-invariant, and further $k$ is necessarily integer-valued and so notions like derivatives with respect to $k$ are not defined.}  Specifically, a smeared probability distribution on dataspace is defined as\footnote{This is effectively a kernel density estimation \cite{rosenblat1956remarks,parzen1962estimation} with a step or window function kernel, but where we are most interested in the variation of the response as a function of width or bin size $\epsilon$.  I thank Rikab Gambhir for identifying this relationship.}
\begin{align}
p(\vec x|\epsilon) \equiv\int d\vec x' p(\vec x')\, \Theta\left(\epsilon - d(\vec x,\vec x')\right)\,,
\end{align}
where $\Theta(x)$ is the Heaviside step function that returns 1 if $x > 0$ and 0 otherwise.  Note that if the original distribution is normalized, then this smeared distribution is not, but this can be adjusted later, if necessary.  What makes this smeared distribution especially powerful is that, for sufficiently large $\epsilon$, even on a discrete and finite dataset, this smeared distribution is well-defined and continuous on the entire dataspace.  As such, ratios of smeared distributions are well-defined everywhere.  Specifically, we can define the smeared likelihood directly as
\begin{align}\label{eq:smearlike}
{\cal L}(\vec x|\epsilon) \equiv \frac{\int d\vec x'\, p_b(\vec x')\, \Theta\left(\epsilon - d(\vec x,\vec x')\right)}{\int d\vec x'\, p_s(\vec x')\, \Theta\left(\epsilon - d(\vec x,\vec x')\right)}\,.
\end{align}
With this smeared likelihood, one can then study its discrimination power on the smeared signal and background distributions as a function of resolution $\epsilon$.  As $\epsilon$ decreases, discrimination power should improve, and resolutions $\epsilon$ at which large jumps in discrimination power occur is thus indicative of a scale of important physics.

However, given a discrete and finite dataset $\{\vec x_i\}_{i=1}^n$, one of course cannot decrease $\epsilon$ arbitrarily because there will be some minimal $\epsilon$ below which there are no events in the neighborhood of other events.  Correspondingly, given a finite dataset size $n$, there is a minimal resolution with which dataspace can be probed, and any scales smaller than that necessarily means that the machine is extrapolating.  In other contexts, it has been empirically observed that there are, rather generally, scaling laws that relate compute resources (like the size of the training dataset) to performance (like the value of the likelihood or objective function), see, e.g., \Refs{ahmad1988scaling,cohn1990can,hestness2017deep,kaplan2020scaling,rosenfeld2019constructive,henighan2020scaling,rosenfeld2021predictability}.  In the present context, the existence of scaling laws between compute resources (like the size of the training dataset) to performance (like the minimal distance over which a machine must extrapolate) follow rather directly as a consequence of extreme value theory \cite{frechet1927loi,fisher1928limiting,von1936distribution,gnedenko1943distribution}.  Events are drawn identically and independently on the dataspace, and the cumulative distribution of metric distances between pairs of events $\Sigma(d)$ is well-defined.  Therefore, extremely generically, on a dataset of $n \to\infty$ events, the (mean) minimal distance between pairs of events $d_n$ will scale like
\begin{align}
n\,\Sigma(d_n)=1\,.
\end{align}
We will show in some well-motivated physics examples, this often implies that $d_n \propto n^\gamma$, at least to good approximation, for some scaling exponent $\gamma$.

In this paper, we will mostly concern ourselves with this general smearing analysis, without restricting to any individual machine learning architecture, but we want to emphasize that this approach nicely applies to understanding the idiosyncratic output of any machine, too.  Let's denote the output of a machine for binary discrimination to be $\hat {\cal L}(\vec x)$, which is an approximation for the functional form of the true likelihood on phase space $\vec x$.  Again, given a typical architecture, the specific way this function is expressed is not interpretable, but we can smear over it to study the energy scales that it exploits.  We define the smeared machine output to be
\begin{align}
\hat {\cal L}(\vec x|\epsilon) \equiv \frac{\int d\vec x'\, p_s(\vec x')\, \hat {\cal L}(\vec x')\, \Theta\left(\epsilon - d(\vec x,\vec x')\right)}{\int d\vec x'\, p_s(\vec x')\, \Theta\left(\epsilon - d(\vec x,\vec x')\right)}\,,
\end{align}
where we note that we have effectively averaged the output over the signal data exclusively.  We use this definition because if $\hat {\cal L}(\vec x)$ is the true likelihood, then this smeared version reduces to the smeared likelihood of \Eq{eq:smearlike}.  We leave a detailed study of this smearing to studying many architectures that are on the market, in a similar way to the approach of \InRef{Geuskens:2024tfo}, to future work.

The outline of this paper is as follows.  In \Sec{sec:semd}, we review the $p=2$ Spectral Energy Mover's Distance metric that will be used throughout the rest of this paper.  In \Sec{sec:qvg}, we study the features of this metric distance smearing through the concrete example of quark versus gluon jet discrimination.  We generate simulated data and study the dependence of minimal resolution on dataset size, and show that discrimination power steadily increases as resolution decreases.  This reflects the (widely-known) property that the likelihood for quark versus gluon jet discrimination is sensitive to emissions at all scales.  We conclude in \Sec{sec:concs} and show how these smeared distributions can be calculated in perturbation theory in the appendix.

\section{Review of Spectral Energy Mover's Distance}\label{sec:semd}

In this section, we will review the particle physics event metric that we will use in the rest of this paper, namely, the $p=2$ Spectral Energy Mover's Distance (SEMD).  This will exclusively be a review of its functional form on the relevant dataspace, and we refer to the original papers for motivation, proofs that it is a metric, efficient algorithms for evaluation, and benchmark tests \cite{Larkoski:2023qnv,Gambhir:2024ndc}.

The $p=2$ Spectral Energy Mover's Distance evaluated between two events ${\cal E}_A, {\cal E}_B$ defined as point-clouds of particles on the celestial sphere is:
\begin{align}
\label{eq:discrete_spectral_emd}
\text{SEMD}_{p=2}({\cal E}_A, {\cal E}_B) &= \sum_{i<j\in{\cal E}_A}2E_iE_j\omega_{ij}^2+\sum_{i<j\in{\cal E}_B}2E_iE_j\omega_{ij}^2-2\sum_{\substack{n\in {\cal E}_A^2,\,\ell\in{\cal E}_B^2\\ \omega_n<\omega_{n+1}\\ \omega_{\ell}<\omega_{\ell+1}}}\omega_n \omega_{\ell}\,\text{ReLU}(\mathcal{S}_{n\ell})\,. 
\end{align}
In this expression, indices $i,j$ label individual particles in the events, $E_i$ is an appropriate energy of particle $i$, and $\omega_{ij}$ is an appropriate angle between the momenta of particles $i$ and $j$.  In the rightmost term, $\text{ReLU}(x) \equiv x \, \Theta(x)$ is the Rectified Linear Unit function~\cite{4082265}.  In this term in particular, $n$ and $\ell$ label a pair of particles from either event $A$ or event $B$, respectively, and pairwise particle angles in each event are ordered (i.e., $\omega_n < \omega_{n+1}$).  The function $S_{n\ell}$ that mixes the events is defined as
\begin{align}
\label{eq:theta_function}
\mathcal{S}_{n\ell} \equiv \min\left[ S_A^+(\omega_n),S_B^+(\omega_{\ell})
\right]-\max\left[ S_A^-(\omega_n),S_B^-(\omega_{\ell})
\right]\,,
\end{align}
where the inclusive and exclusive cumulative spectral functions are:
\begin{align}
S^+(\omega_n) & = \sum_{i\in {\cal E}} E_i^2+\sum_{\substack{n\geq m\in{\cal E}^2\\\omega_m<\omega_{m+1}}}(2EE)_m\,,\\
S^-(\omega_n) &= \sum_{i\in {\cal E}} E_i^2+\sum_{\substack{n> m\in{\cal E}^2\\\omega_m<\omega_{m+1}}}(2EE)_m\,.
\end{align}
Finally, the shorthand $(2EE)_m = 2E_iE_j$, for particles $i$ and $j$, assuming that the pair $(i,j) = m$.

This expression for the SEMD actually produces a squared metric distance, and so the true metric (that satisfies the triangle inequality) is its square-root:
\begin{align}
d({\cal E}_A,{\cal E}_B) = \sqrt{\text{SEMD}_{p=2}({\cal E}_A, {\cal E}_B)}\,.
\end{align}
In this paper, we will restrict our analysis to events or individual jets at a hadron collider, and so will use appropriate coordinates for such events.  Energies therefore will be measured by transverse momentum to the collision beam, $E_i \to p_{\perp,i}$, and pairwise angles as distances in the rapidity-azimuth plane $(y,\phi)$, where
\begin{align}
    \label{eq:metric_patch}
    \omega_{ij}^2 = (y_i - y_j)^2+(\phi_i - \phi_j)^2\,.
\end{align}
For all results that follow, we used the code SPECTER to evaluate the SEMD, which can be downloaded from \url{https://github.com/rikab/SPECTER}.

While we use the $p=2$ SEMD exclusively in this paper, other IRC safe metrics could be applied to the same analysis and the differences between them would be interesting.  Such a study would shine some light on the space of IRC safe metrics and the different physics each are more or less sensitive to, much in the same way as distinct IRC safe jet algorithms cluster radiation differently in an event.  However, as of now, there exist no other IRC safe metrics that have been proposed that can both be expressed exactly in closed form as a function of coordinates on phase space (requiring no additional minimization procedure) and can be evaluated fast enough to be applied on a large dataset with finite compute resources.

\section{Case Study: Quark versus Gluon Jet Discrimination}\label{sec:qvg}

As a concrete testing ground for this smearing procedure, we will study the problem of identification and discrimination of jets initiated by light quarks versus gluons.  This is an ancient problem in collider physics \cite{Jones:1988ay,Fodor:1989ir,Lonnblad:1990bi,Lonnblad:1990qp,Csabai:1990tg,Jones:1990rz,Pumplin:1991kc,OPAL:1993uun}, with some of the first neural networks applied to particle physics analyses used for this purpose.  It has long been known that simply the particle multiplicity is a very powerful discrimination observable itself, and further in the double logarithmic approximation, is in fact (monotonically related to) the likelihood \cite{benchat,Frye:2017yrw,Bright-Thonney:2022xkx}.  Inclusive jet production has only one large energy scale imposed, the scale of the total jet's energy, and approximate scale invariance of particle production in quantum chromodynamics (QCD) suggests that sensitivity to physics at all scales is necessary for optimal quark versus gluon discrimination.  Indeed, that total particle multiplicity is a powerful observable is indicative of this expectation.

\subsection{Event Generation and Analysis}

Our quark- and gluon-initiated jet samples are generated first at leading-order from $pp\to q(Z\to \nu\bar \nu)$ and $pp\to g(Z\to \nu\bar \nu)$ events in MadGraph5 v3.6.0 \cite{Alwall:2014hca} at the 13 TeV Large Hadron Collider.  The events were then showered and hadronized with default settings with Pythia v8.306 \cite{Bierlich:2022pfr}.  All particles except neutrinos were recorded for further analysis.  Anti-$k_T$ \cite{Cacciari:2008gp} jets with radius $R = 0.5$ were clustered with FastJet v3.4.0 \cite{Cacciari:2011ma}, and we only kept the leading jet with transverse momentum in the range $500 < p_\perp < 550$ GeV and pseudorapidity $|\eta| < 2.5$.  Only a maximum of 100 particles in each jet was recorded for further analysis; if a jet contained more than 100 particles, the jet was reclustered with the exclusive $k_T$ algorithm \cite{Catani:1993hr,Ellis:1993tq} down to 100 particles.  Limiting to 100 particles per jet was a very weak restriction.  Gluon jets in this sample had a mean multiplicity of about 59 particles, and a standard deviation of about $\sigma_g = 17.6$, and so 100 particles was more than $2\sigma_g$ from the mean.  On quark jets, the mean multiplicity was about 38 particles, with a standard deviation of about $\sigma_q = 15.2$, and so 100 particles was more than $4\sigma_q$ away.

This definition of quark and gluon jets from the leading-order event selection has a long tradition in jet physics \cite{Gras:2017jty}, but is not technically theoretically well-defined.  Recently, several infrared (and collinear) safe definitions of the flavor of a jet have been proposed \cite{Banfi:2006hf,Caletti:2022hnc,Caletti:2022glq,Czakon:2022wam,Gauld:2022lem,Caola:2023wpj}, which all necessarily reduce to the naive leading-order selection, but in general differ at higher orders.  While a robust flavor definition is necessary to draw quantitative conclusions from data, we stick to the simple, practical definition of ``quark'' and ``gluon'' as the output of simulation, because of its ubiquity, but otherwise apologize for further perpetuating this imprecision.

Then, on these showered, hadronized, and, if necessary, reclustered, jets, we evaluated the SEMD between all pairs of jets with SPECTER.  We used an A100 GPU with 40 GB of GPU RAM and 83.5 GB of System RAM, and processed pairs of events in batch sizes of 10000.  We evaluated all pairwise distances between quark and gluon jet samples of 20000 events each, for a total of 799980000 unique distances to calculate.  On average, each pairwise SEMD took about $1.3\times 10^{-5}$ seconds per evaluation.  Finally, all pairwise distances were recorded for further analysis.

\subsection{Minimal Smearing versus Dataset Size}

The first thing we need to establish is what the minimal smearing resolution scale $\epsilon_{\min}$ is such that this is still meaningful.  Recall the expression for the smeared likelihood for this problem
\begin{align}
{\cal L}(\vec x|\epsilon) = \frac{\int d\vec x'\, p_q(\vec x')\, \Theta\left(\epsilon - d(\vec x,\vec x')\right)}{\int d\vec x'\, p_g(\vec x')\, \Theta\left(\epsilon - d(\vec x,\vec x')\right)}\,.
\end{align}
On our event ensembles, the phase space points $\vec x$ at which we evaluate the smeared likelihood will be points where we have events.  An event of a given class is always a distance 0 from itself, and so ``same class'' smeared distributions will necessarily be non-trivially bounded from below.  That is, for a gluon event located at phase space point $\vec x_g$, say, we have
\begin{align}
\int d\vec x'\, p_g(\vec x')\, \Theta\left(\epsilon - d(\vec x_g,\vec x')\right)\geq \frac{1}{n}\,,
\end{align}
for a dataset of a total of $n$ events.  For the likelihood to be useful, it can't simply evaluate to 0 or $\infty$, which enforces a limit through the distinct class smearing.  For example, a quark event at phase space point $\vec x_q$ can possibly have 0 gluon events within $\epsilon$:
\begin{align}
\int d\vec x'\, p_g(\vec x')\, \Theta\left(\epsilon - d(\vec x_q,\vec x')\right)\geq 0\,.
\end{align}
However, there will be a minimal distance $\epsilon_{\min}$ at which this is at least $1/n$:
\begin{align}
\int d\vec x'\, p_g(\vec x')\, \Theta\left(\epsilon_{\min} - d(\vec x_q,\vec x')\right)\geq \frac{1}{n}\,.
\end{align}
We evaluate the minimal distance between a gluon event and any quark event, and between a quark event and any gluon event, and can correspondingly interpret the resulting distributions.

\subsubsection{Empirical Observations}

\begin{figure}[t!]
\begin{center}
\includegraphics[width=0.45\textwidth]{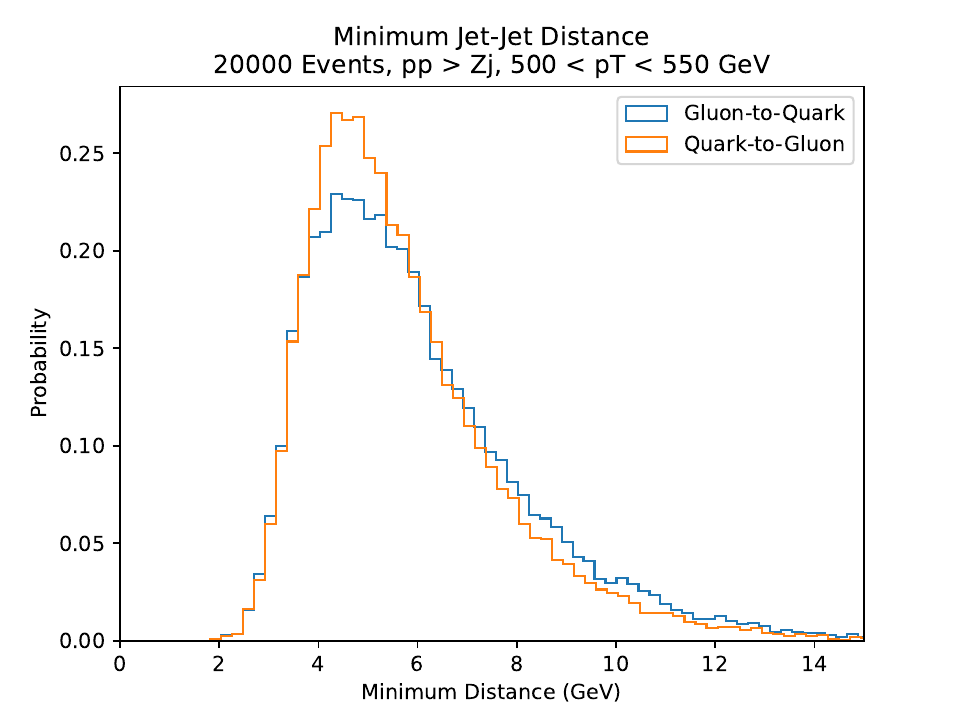}\ \ \ 
\includegraphics[width=0.45\textwidth]{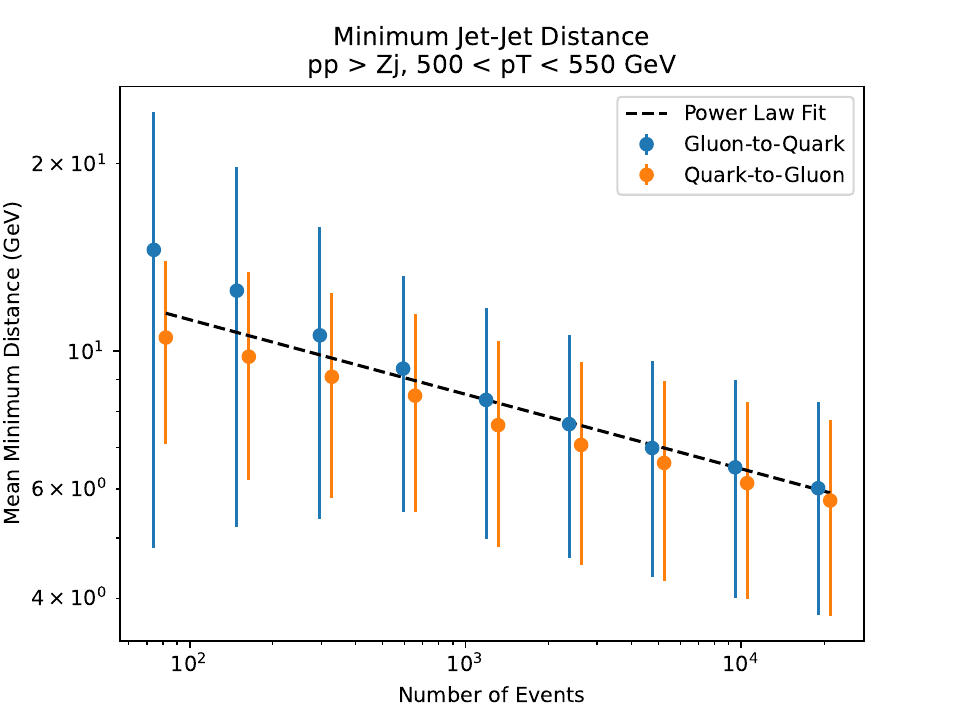}
\caption{\label{fig:exvalth}
Left: distribution of the minimal distances in GeV between gluon jets and any quark jets (blue), and between quark jets and any gluon jets (orange), on the 20000+20000 jet dataset.  Right: Plot of the relationship between the mean minimum distance $\langle\epsilon_{\min}\rangle$ in GeV on gluon-to-quark (blue) and quark-to-gluon (orange) jet distances, as a function of number of events $n$ in the dataset.  Points are displaced $\pm 5\%$ from the true number of events for visibility, and the vertical line represents $\pm 1$ standard deviation about the means.  The power law fit of $\langle \epsilon_\text{min}\rangle = 19.4\,n^{-0.12}$ GeV is shown in dashed-black.
}
\end{center}
\end{figure}

At left in \Fig{fig:exvalth}, we plot the distribution of these minimal gluon-quark (blue) and quark-gluon (orange) jet-jet distances on the complete 20000+20000 event dataset.  This demonstrates that the minimum meaningful smearing $\epsilon$ that can be used is about $\epsilon = 10$ GeV, at which only a few percent of the events will have a likelihood directly evaluate to 0 or infinity.  One thing in particular to note is that a scale of 10 GeV is perturbative, suggesting that smearing at this scale will correspond to summing over jets that differ by perturbative emissions.  From our expectations discussed earlier, we would want the likelihood to be sensitive to all emissions in the jet, which means smearing over jets that only differ by sufficiently low-scale or non-perturbative emissions, below a scale of about 1 GeV.\footnote{Recall that the definition of the SEMD is closely related to the sum of the squared masses of the jets.  As such, a 10 GeV contribution to the jet mass can come from a wide-angle non-perturbative emission that has a relative transverse momentum of order of 1 GeV, if the jet mass is already relatively small.  Here, we are considering sensitivity to the emission of one more or one fewer hadron in a jet, or particular sensitivity to the precise value of hadron masses.  On this latter point, it may be preferred to modify the definition of the SEMD to include hadron masses explicitly, rather than to just use transverse momentum.  Related studies on the effect of hadron masses on IRC safe observables are presented in \Refs{Salam:2001bd,Mateu:2012nk}.}  Apparently this sample of 20000+20000 jets does not enable this, but what dataset size does?

To answer this question, at right in \Fig{fig:exvalth}, we plot the mean minimum distance as a function of number of jets in the dataset.  We also display $\pm 1\sigma$ on this plot and the quark-gluon and gluon-quark values are displaced by 5\% above or below the exact number of events studied for visibility.  On this log-log plot, the dependence of the means on the number of events is approximately linear, indicative of a power-law relationship.  From a naive fit, the mean minimum distance for gluon-to-quark jets approximately satisfies the scaling law
\begin{align}\label{eq:approxscale}
\langle \epsilon_\text{min}\rangle \approx 19.4 \,n^{-0.12}\text{ GeV}\,,
\end{align}
where $n$ is the number of jets in the sample.  For this mean to be comparable to the scale of individual hadron emissions, $\langle \epsilon_\text{min}\rangle\sim 1$ GeV, the datasize would have to exceed $10^{10}$ events, which is still currently several orders of magnitude larger than even the largest datasets used for training in particle physics applications, e.g., \Refs{Qu:2022mxj,Amram:2024fjg}.  This simple observation demonstrates that on any practical dataset on which a machine for quark versus gluon jet discrimination is trained, that machine necessarily must extrapolate between events separated by emissions at perturbative energy scales.  This may suggest that ensuring that the machine explicitly knows non-perturbative information like the total multiplicity may be useful in reducing the extrapolation distance, even given a limited training set size.

\subsubsection{Extreme Value Theory Analysis}

For an analytic understanding of this scaling behavior, we consider a closely related, but somewhat simpler, problem than what we study above.  We study the distribution of the minimum distance $d_{\min}$ on a random sample of $n$ pairs of quark-gluon jets.  All events of a given class are drawn independently and from the same distribution, and so this problem satisfies the assumptions of the Fisher-Tippett-Gnedenko theorem \cite{frechet1927loi,fisher1928limiting,von1936distribution,gnedenko1943distribution}, and so extreme value theory can be applied.  This can be used to predict the effective scaling exponent of the mean minimum distance, given the appropriate distribution of pairwise event distances.  We will demonstrate how this works to lowest meaningful perturbative order for these quark and gluon jets.

Note that the SEMD is proportional to the invariant mass $s$ of a jet, at lowest order, $d^2 \propto s$.  To double logarithmic accuracy, we know the cumulative distribution of the invariant mass as a Sudakov factor, and to determine the distance distribution requires appropriately adjusting color factors to account for the fact that we are studying the distance between jets in different event classes.  The cumulative distribution of the distance between quark and gluon jets at double logarithmic accuracy is \cite{Komiske:2022vxg,Larkoski:2023qnv}
\begin{align}
\Sigma(d) = \exp\left(-\frac{2\alpha_s (C_A+C_F)}{\pi}\log^2\frac{d}{E}\right)\,,
\end{align}
where $E$ is the jet energy, $C_F = 4/3$ is the fundamental and $C_A=3$ is the adjoint Casimir of SU(3) color.  We want the distribution of the minimum value with $n$ draws from this distribution, so we need to analyze the behavior of $1-\Sigma(d)$ to the $n$th power, as $n\to\infty$.  The probability that all $n$ events have minimal distances larger than some $d$ is
\begin{align}
\lim_{n\to\infty}\left(1-\Sigma(d)\right)^n = \exp\left(
-n\,\Sigma(d)
\right)=\exp\left(
-n \exp\left(-\frac{2\alpha_s (C_A+C_F)}{\pi}\log^2\frac{d}{E}\right)
\right)\,.
\end{align}
Now, we define $d_n$ as
\begin{align}
n\, \Sigma(d_n) = n \exp\left(-\frac{2\alpha_s (C_A+C_F)}{\pi}\log^2\frac{d_n}{E}\right) = 1\,,
\end{align}
or, that,
\begin{align}\label{eq:minval}
d_n = E \exp\left(
-\sqrt{\frac{\pi \log n}{2\alpha_s (C_A+C_F)}}
\right)\,.
\end{align}
Expanding the exponent about this value produces the cumulative distribution of the minimal value of the distance with $n$ events, where
\begin{align}
\Sigma(d_\text{min}|n) =1- \exp\left(
-\exp\left(
\frac{d_{\min}-E\,e^{-\sqrt{\frac{\pi\log n}{2\alpha_s (C_A+C_F)}}}}{E\, \sqrt{\frac{\pi}{8\alpha_s (C_A+C_F)\log n}}\,e^{-\sqrt{\frac{\pi\log n}{2\alpha_s (C_A+C_F)}}}}
\right)
\right)\,.
\end{align}
This is effectively a Gumbel distribution \cite{gumbel1935valeurs,gumbel1941return}, modified from the usual distribution of maxima to distribution of minima.\footnote{An additional property of this distribution to note is that its coefficient of variation, or ratio of standard deviation to mean, is approximately constant, and only weakly dependent on number of events $n$, for a large range of $n$.  This can be observed through the ratio of the scale factor, the denominator of the exponent, to the central value, the subtracted factor in the numerator of the exponent, which serve as proxies for the standard deviation and mean, respectively.  Measurements that are log-normally distributed exhibit a stationary coefficient of variation, and this feature of this minimum distance distribution may be a consequence of the near log-normal form of the Sudakov factor, but it would be interesting to study if this property arises for distance distributions that are not approximately log-normal.  I thank Yoni Kahn for this observation.}

From this distribution, the mean can be calculated, and then the corresponding scaling with number of events $n$ can be found.  However, it is much simpler to just use the expression of \Eq{eq:minval}, which will exhibit the same scaling as $n\to \infty$ as the mean.  If we assumed that $d_n$ took a power-law form, where
\begin{align}
d_n = d_0 n^{\gamma}\,,
\end{align}
where $d_0$ is some reference distance and $\gamma$ is the scaling dimension, then we can extract the scaling dimension via
\begin{align}
\frac{n}{d_n} \frac{d}{dn}\,d_n = \gamma = -\sqrt{\frac{\pi}{8\alpha_s (C_A+C_F) \log n}}\,.
\end{align}
This is not independent of number of events $n$, and so is not a true scaling dimension.  However, the residual $n$ dependence varies extremely slowly, and so in a plot of just a few orders-of-magnitude, it is unlikely that a deviation would be observed easily.  We also note that the value of the effective scaling exponent $\gamma$ in this double logarithmic approximation, and what we empirically observe in \Eq{eq:approxscale}, is approximately the same size as scaling exponents extracted in other discrimination observable settings \cite{Batson:2023ohn}.  More detailed studies will be needed to establish the theory behind more general scaling exponents, but this is a concrete and promising approach.

This extreme value theory analysis is suggestive of a general procedure for predicting scaling exponents.  Given that we can calculate the cumulative distribution of distances $\Sigma(d)$, then the characteristic distance $d_n$ on a dataset of $n$ events is defined implicitly via
\begin{align}
n\,\Sigma(d_n) = 1\,.
\end{align}
We note that scaling laws are only expected to hold in the asymptotic $n\to \infty$ limit, and so to evaluate $d_n$, we only need the cumulative distribution in the small-distance limit, $\lim_{d\to 0}\Sigma(d)$.  For the quark-to-gluon jet distance at hand, in this limit, the SEMD is proportional to the squared jet mass, and the squared jet mass is IRC safe and additive, whereby each additional emission in the jet contributes a strictly positive amount that adds to the jet mass.  As such, the cumulative distribution of distances takes a general form where \cite{Catani:1992ua}
\begin{align}
\lim_{d\to 0}\Sigma(d)  = C(\alpha_s)\exp\left[
Lg_0(\alpha_s L)+g_1(\alpha_s L)+\alpha_s g_2(\alpha_s L)+\cdots
\right]\,,
\end{align}
where $C(\alpha_s)$ is some function of the strong coupling $\alpha_s$, $L = \log(d/E)$, and the $g_i(x)$ are functions of $x = \alpha_s L$.  These functions can be calculated at fixed-order and including through $g_k(x)$ in the exponent is resummation through (next-to-)$^k$ leading logarithmic accuracy.  Even the full leading-logarithmic function $g_0(\alpha_s L)$ cannot be inverted with elementary functions to express $d_n$ in an interpretable form, which is why we stuck to double logarithmic accuracy above.  However, this inversion can of course be done numerically, which can correspondingly improve the prediction of the scaling exponent, and its deviation from scale invariant.

\subsubsection{Analysis on Jets with Explicit Scales}

For different discrimination problems, this extreme value theory analysis will be different, because the physics that governs the distribution of pairwise event distances is different.  For example, consider the problem of discrimination of massive QCD jets from boosted, hadronic decays of massive particles, such as $W$, $Z$, or Higgs bosons.  This problem is similarly ancient within the field of jet substructure \cite{Seymour:1993mx,Butterworth:2002tt,Butterworth:2007ke,Butterworth:2008iy}, and had particular historical import with reigniting interest and feasibility of finding $H\to b\bar b$ decays.  At any rate, what is especially important and distinct from the quark versus gluon problem is that one imposes an explicit mass $m_J$ on purported jets of interest, and then searches for further correlations amongst emissions inside of them.  We can correspondingly expand the signal and background distributions with a fixed jet mass in powers of the strong coupling $\alpha_s$ as:
\begin{align}
p_s(\vec x|m_J) = p_s^{(0)}(\vec x|m_J) +\frac{\alpha_s}{2\pi}\, p_s^{(1)}(\vec x|m_J) +\cdots\,,\\
p_b(\vec x|m_J) = p_b^{(0)}(\vec x|m_J) +\frac{\alpha_s}{2\pi}\, p_b^{(1)}(\vec x|m_J) +\cdots\,,
\end{align}
where the superscript $(i)$ denotes the term at order $\alpha_s^i$.  Because of the mass constraint, the leading-order distributions $p^{(0)}(\vec x|m_J)$ are themselves necessarily positive and normalizable, and so are probability distributions.

Then, we can directly calculate the distribution of pairwise distances between signal and background events, expanded in powers of $\alpha_s$.  The cumulative distribution of pairwise distances $d$ is then
\begin{align}
\Sigma(d) &= \int d\vec x\, d\vec x'\, p_s(\vec x|m_J)\, p_b(\vec x'|m_J)\, \Theta\left(d - d(\vec x,\vec x')\right)\\
&=\int d\vec x\, d\vec x'\, p^{(0)}_s(\vec x|m_J)\, p^{(0)}_b(\vec x'|m_J)\, \Theta\left(d - d(\vec x,\vec x')\right)\nonumber\\
&\hspace{1cm}+\frac{\alpha_s}{2\pi}\int d\vec x\, d\vec x'\left[ p^{(1)}_s(\vec x|m_J)\, p^{(0)}_b(\vec x'|m_J)+p^{(0)}_s(\vec x|m_J)\, p^{(1)}_b(\vec x'|m_J)\right] \Theta\left(d - d(\vec x,\vec x')\right)+\cdots\,,\nonumber
\end{align}
writing out the first two orders explicitly.  For extreme value theory analysis, we only need the $d\to 0$ limiting behavior of this distribution, so we can Taylor expand the distribution at leading-order to produce
\begin{align}
\lim_{d\to 0}\Sigma(d) &=d \int d\vec x\, d\vec x'\, p^{(0)}_s(\vec x|m_J)\, p^{(0)}_b(\vec x'|m_J)\, \delta\left( d(\vec x,\vec x')\right)\\
&\hspace{2cm}-\frac{d^2}{2} \int d\vec x\, d\vec x'\, p^{(0)}_s(\vec x|m_J)\, p^{(0)}_b(\vec x'|m_J)\, \delta'\left( d(\vec x,\vec x')\right)+{\cal O}(\alpha_s)\,.\nonumber
\end{align}
By smoothness, positivity, and monotonicity, the cumulative distribution must vanish as $d\to 0$ and so no constant term appears.  

In this expression, we have expanded the cumulative distribution to quadratic order in $d$ because the linear term actually vanishes, for the $p=2$ SEMD metric that we consider in this paper.  To demonstrate this, we note that the metric between two jets each with total mass $m_J$ can be expressed as
\begin{align}
d(\vec x,\vec x')^2 = \text{SEMD}_{p=2}({\cal E}_A, {\cal E}_B) &= 4m_J^2-2\sum_{\substack{n\in {\cal E}_A^2,\,\ell\in{\cal E}_B^2\\ \omega_n<\omega_{n+1}\\ \omega_{\ell}<\omega_{\ell+1}}}\omega_n \omega_{\ell}\,\text{ReLU}(\mathcal{S}_{n\ell})\,,
\end{align}
where, additionally, we note that we work in the collinear limit.  Now, we note that in this expression the squared metric is linear in pairwise angles $\omega_n$, for example, so we can evaluate the integral over one pairwise angle.  In these coordinates, the $\delta$-function of the metric takes the general form
\begin{align}
\int d\vec x \, \delta\left(d(\vec x,\vec x')\right) \supset \int d\omega\,  \delta\left(\sqrt{{\cal S}_0-{\cal S}_\omega \omega}\right)\,,
\end{align}
where ${\cal S}_0$ is the contribution to the metric that is independent of the angle of interest $\omega$, and ${\cal S}_\omega$ is the coefficient of the term proportional to the angle $\omega$.  This integral can then be evaluated and we find
\begin{align}
\int d\omega\,  \delta\left(\sqrt{{\cal S}_0-{\cal S}_\omega \omega}\right) = \int d\omega \,\frac{2}{{\cal S}_\omega}\,\sqrt{{\cal S}_0-{\cal S}_\omega \omega}\,\delta\left(
\omega - \frac{{\cal S}_0}{{\cal S}_\omega}
\right) = 0\,,
\end{align}
where we have assumed that the probability distribution is finite, smooth, and non-zero at this otherwise not special value of $\omega$.

Thus, the term linear in the distance $d$ vanishes in the cumulative distribution, and in general, with our choice of metric, the cumulative distribution scales quadratically in the distance $d$, as $d\to 0$:
\begin{align}\label{eq:lodist_scaling}
\lim_{d\to 0}\Sigma(d) &=-\frac{d^2}{2} \int d\vec x\, d\vec x'\, p^{(0)}_s(\vec x|m_J)\, p^{(0)}_b(\vec x'|m_J)\, \delta'\left( d(\vec x,\vec x')\right)+{\cal O}(\alpha_s)\,.
\end{align}
This then implies the relationship between the number of events in the dataset $n$ and the characteristic minimal distance $d_n$ of
\begin{align}
n\, \Sigma(d_n)=1 \propto n d_n^2\left(1 +{\cal O}(\alpha_s)\right)\,.
\end{align}
That is, the mean minimal distance for this problem scales with number of events $n$ like
\begin{align}
d_n \propto n^{-1/2+{\cal O}(\alpha_s)}\,.
\end{align}
That is, the scaling exponent $\gamma = -1/2+{\cal O}(\alpha_s)$, where the logarithms that arise from higher-order corrections modify the scaling exponent value away from $-1/2$.\footnote{A closer study of the leading-order term explicit in \Eq{eq:lodist_scaling}, with a derivative of a $\delta$-function, shows that this integral actually does not exist.  This suggests that already at leading order there are logarithms that modify the $d_n\propto n^{-1/2}$ scaling relation, but they may also be non-universal, and depend on the dimensionality of leading-order phase space, for example.  We leave a detailed study with a complete calculation of scaling behavior in concrete examples of discrimination of massive jets to future work.}  This scaling may also be related to similar resolution-limited scaling studied and predicted in the context of large language models, e.g., \Refs{DBLP:journals/corr/abs-2105-01867,bahri2024explaining}, but we leave study of a deeper connection to future work.

\subsection{Discrimination Performance}

\begin{figure}[t!]
\begin{center}
\includegraphics[width=0.45\textwidth]{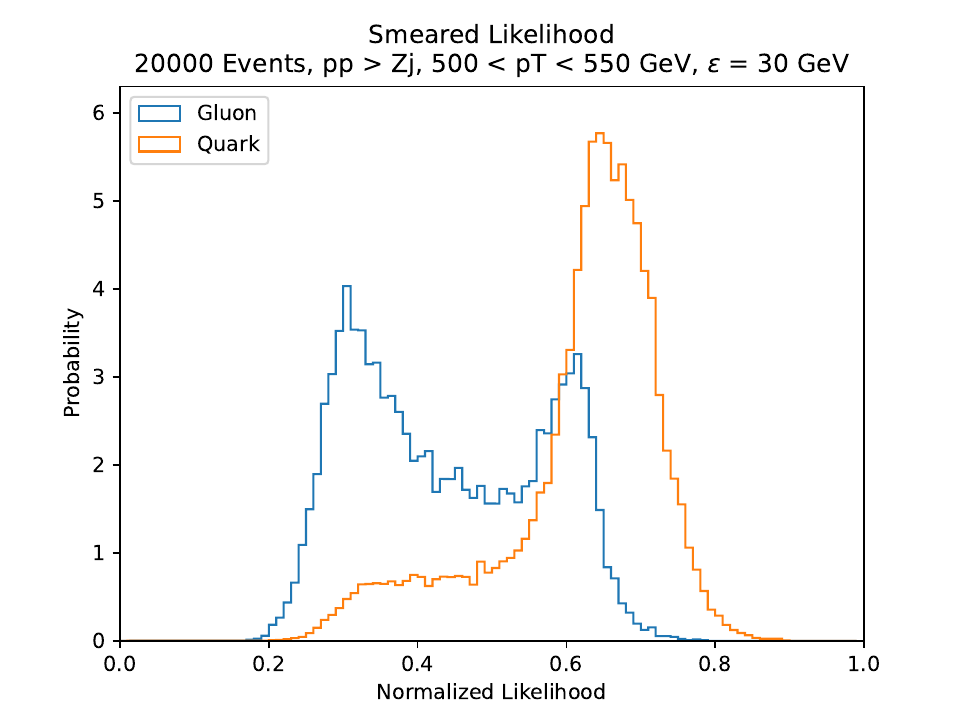} \ \ \ 
\includegraphics[width=0.45\textwidth]{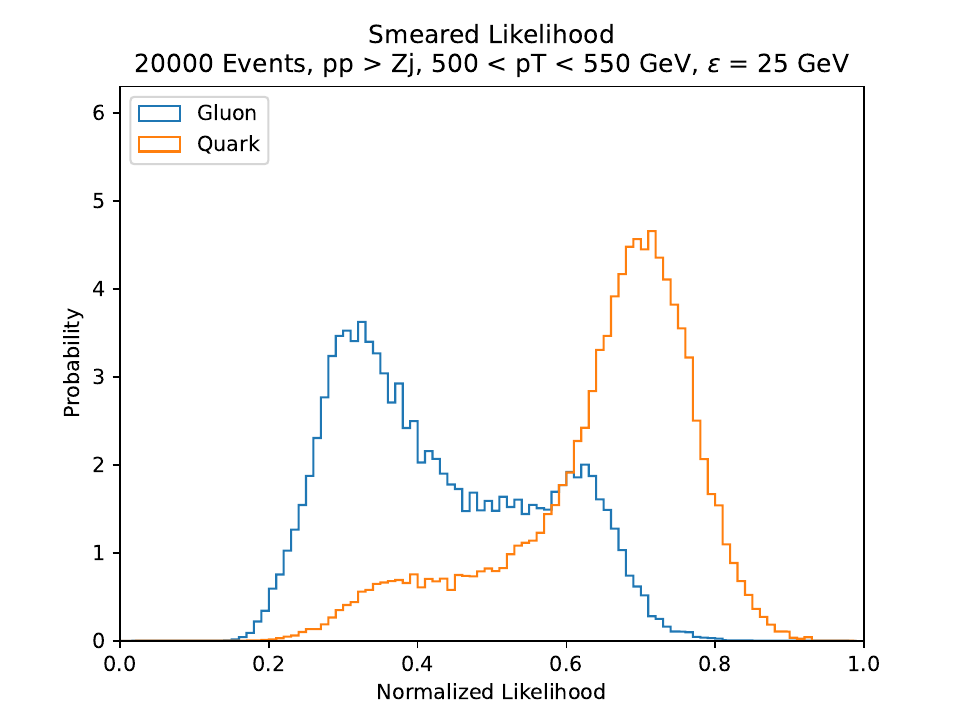}\\
\includegraphics[width=0.45\textwidth]{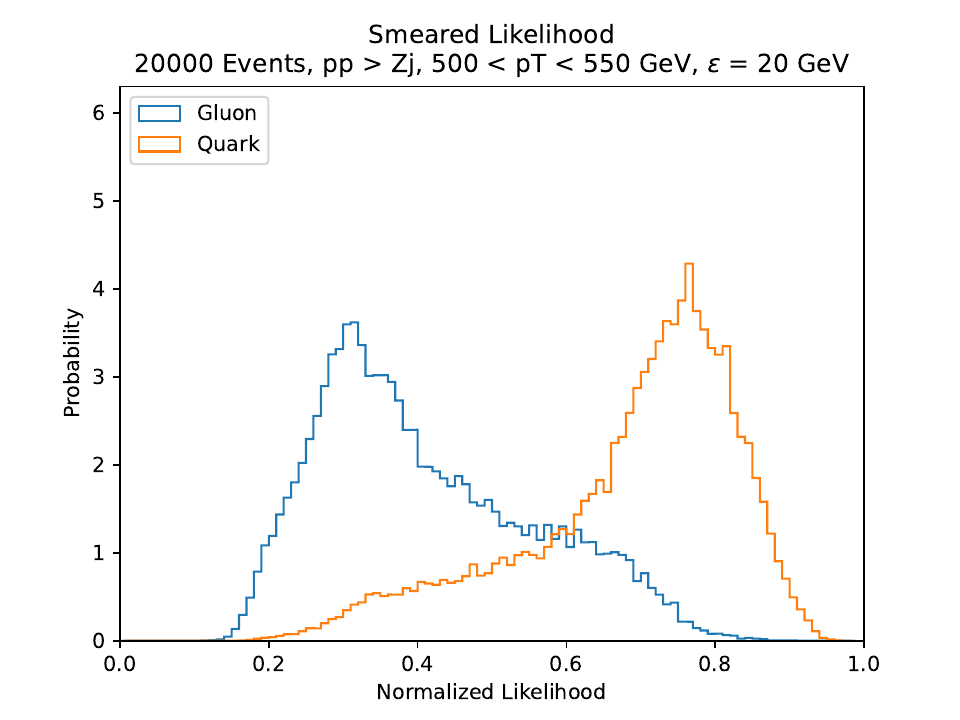} \ \ \ 
\includegraphics[width=0.45\textwidth]{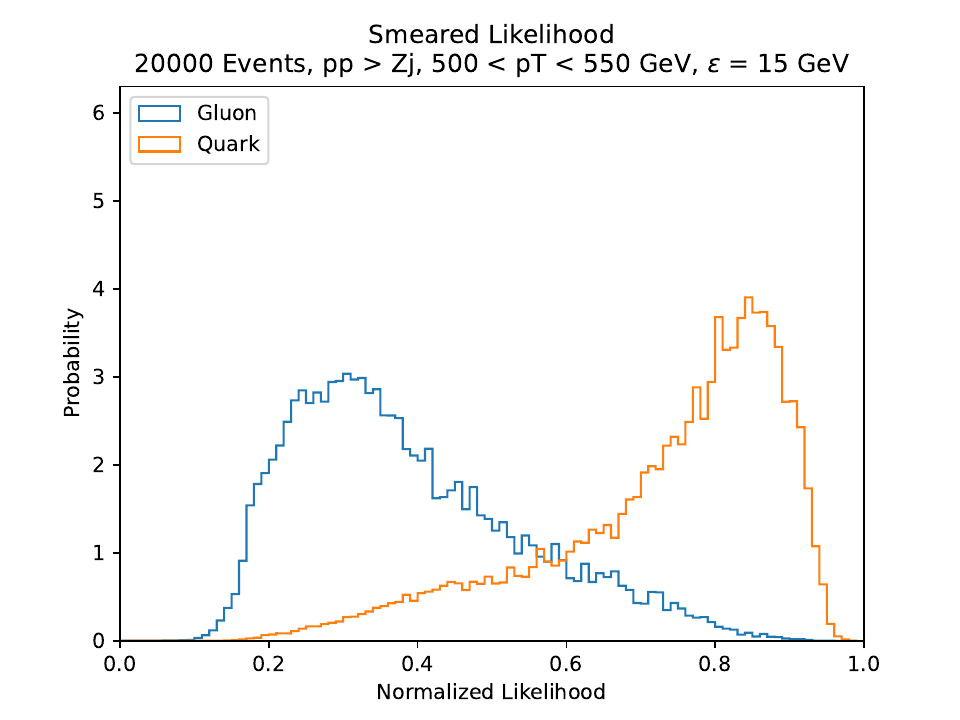}\\
\includegraphics[width=0.45\textwidth]{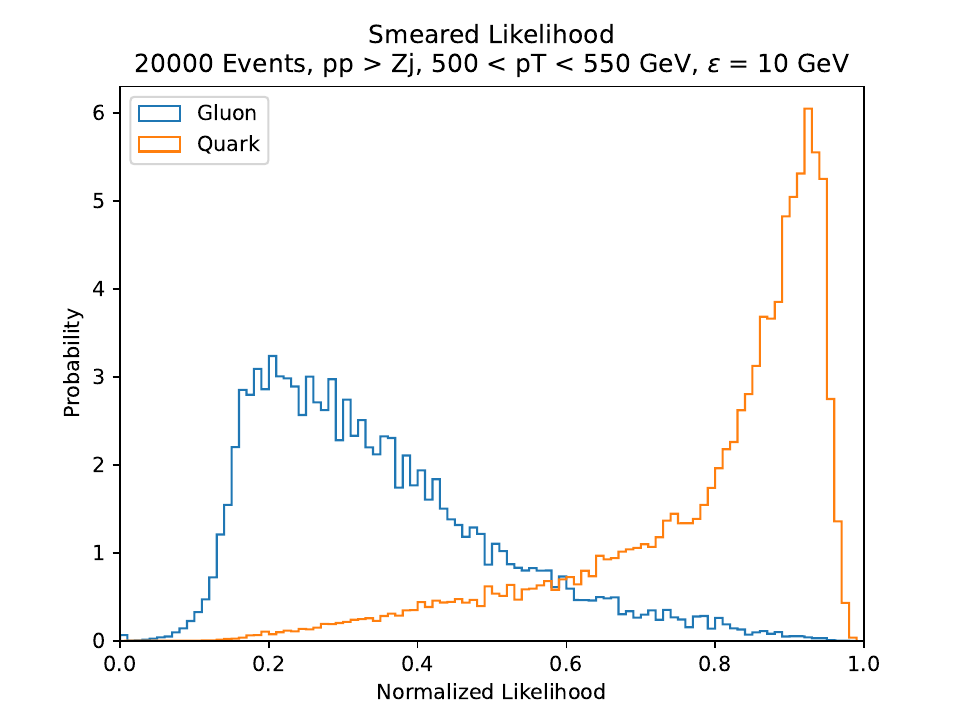}
\caption{\label{fig:likedists}
Distributions of the smeared likelihood score on 20000 events each of gluon jets (blue) and quark jets (orange).  From top left to bottom, the smearing distance $\epsilon$ is varied on $30,25,20,15,10$ GeV.
}
\end{center}
\end{figure}

With this smeared distribution and likelihood analysis, we can then study the discrimination power performance as a function of the smearing resolution $\epsilon$.  From the minimal distance distribution analysis of the previous section, with the 20000+20000 event sample, we can only consider smearing distances down to about $\epsilon = 10$ GeV, and here we will explicitly consider $\epsilon = 10,15,20,25,30$ GeV, to observe some dependence on discrimination power as $\epsilon$ decreases.  To more easily determine how signal and background smeared likelihood distributions shift as a function of $\epsilon$, we will actually plot a normalized version of the smeared likelihood, or a smeared likelihood score ${\cal S}(\vec x|\epsilon)$, where
\begin{align}
{\cal S}(\vec x|\epsilon) = \frac{{\cal L}(\vec x|\epsilon)}{1+{\cal L}(\vec x|\epsilon)} = \frac{p_q(\vec x|\epsilon)}{p_q(\vec x|\epsilon)+p_g(\vec x|\epsilon)}\,.
\end{align}
This is monotonically related to the smeared likelihood, so has the same discrimination power, but is bounded on ${\cal S}(\vec x|\epsilon) \in[0,1]$.

In \Fig{fig:likedists}, we plot the distribution of this normalized likelihood on the quark and gluon samples, for the different smearing distances $\epsilon$ listed earlier.  These plots clearly illustrate less overlap (and therefore, better discrimination), as the smearing scale $\epsilon$ decreases.  Further, at sufficiently large values of $\epsilon$, for about $\epsilon > 25$ GeV or so, the distributions have a double humped structure, indicative of two different populations of events that are being smeared over.  This may be indicative of especially gluon-initiated jets faking quark-initiated jets through a $g\to q\bar q$ splitting that occurs at relatively high scales in the parton shower.  However, we emphasize that we currently do not have a good understanding of the origin of this structure, and look toward a better understanding.  This structure vanishes at smaller smearing distances, and the distributions become strongly peaked at opposite ends of the range of the observable, as expected.


\begin{figure}[t!]
\begin{center}
\includegraphics[width=0.6\textwidth]{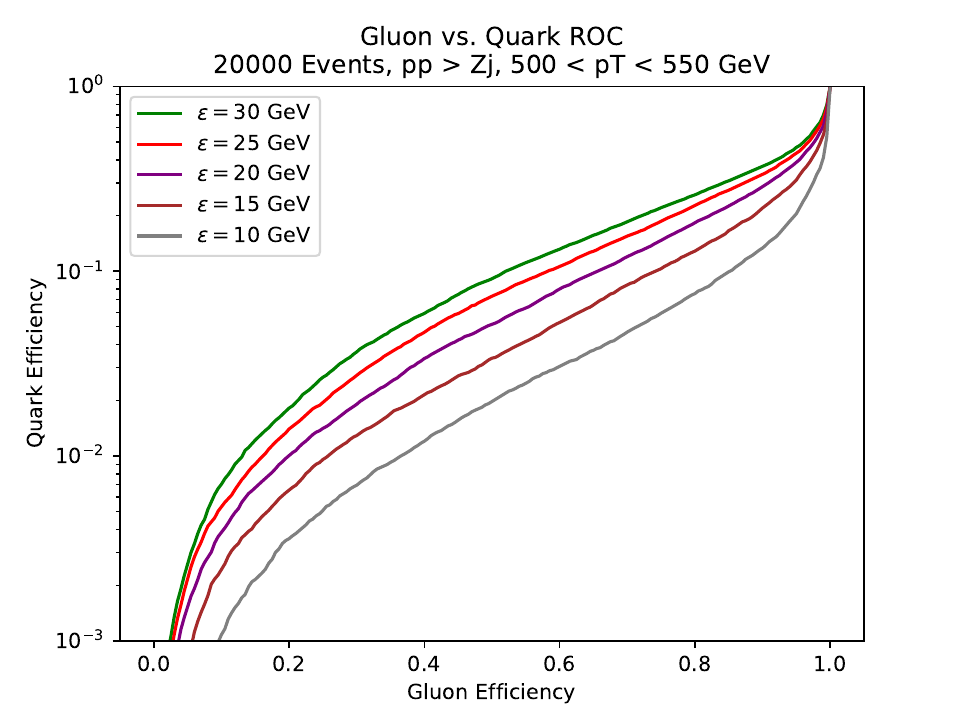}
\caption{\label{fig:roccurve}
ROC curves of quark jet efficiency as a function of gluon jet efficiency from a sliding cut on the smeared likelihood.  The smearing distance is varied from $\epsilon = 30,25,20,15,10$ GeV from top curve to bottom.  Better discrimination performance is the lower right direction of this plot.
}
\end{center}
\end{figure}

From these likelihood distributions, we then plot the corresponding receiver operating characteristic (ROC) curves in \Fig{fig:roccurve}.  As the smearing distance $\epsilon$ decreases, the discrimination power smoothly improves, as indicated by reduced quark efficiency at fixed gluon efficiency.  The smoothness of these curves with smearing distance $\epsilon$ also indicates, at least on this jet sample, that there is no physics at a fixed energy scale that is especially important for discrimination; or that, quark versus gluon discrimination is approximately scale invariant.   Another aspect of these curves to note is that at large gluon efficiencies, the slope of the ROC curve appears to diverge, meaning that the quark efficiency increases by a large amount, while the gluon efficiency changes only slightly.  This is further indicative of the known fact that there exist regions of phase space in which a pure sample of quark jets can be isolated, with no gluon jet contamination \cite{Metodiev:2018ftz}.  

\section{Conclusions}\label{sec:concs}

If we are to claim understanding of what a machine is learning especially in the realm of particle physics, we must confront and solve the problem of interpretability.  
In this paper, we present a first study of smearing over the training data, which requires a metric on the space of events and through it the physics at different distance scales can be studied.  For smearing over sufficiently large distances, this method renders discrete, finite datasets continuous over the entire dataspace, and so meaningful ratios of distributions can be taken, for example, to define appropriately smeared likelihood ratios or smeared machine outputs.  Through extreme value theory, the relationship between the minimal smearing distance and the number of events in the dataset can be defined and calculated, and in many cases produce (approximate) power-law relationships that have been empirically observed in other machine learning contexts, e.g., \Refs{ahmad1988scaling,cohn1990can,hestness2017deep,kaplan2020scaling,rosenfeld2019constructive,henighan2020scaling,rosenfeld2021predictability,Batson:2023ohn}.

The analysis presented here is merely the first glimpse of the utility of this approach to interpretability that we believe can bear much fruit.  This smeared likelihood analysis can be applied to many other discrimination problems in particle physics, such as discrimination of hadronic decays of electroweak bosons or top quarks from QCD jets.  The use of an IRC safe metric and simplicity of the smearing prescription means that first principles calculations can be performed and important physical scales predicted before they are observed in (simulated) data.  One aspect that would be particularly interesting to understand and predict of this analysis is a renormalization group-like flow of the smeared likelihood (or any smeared observable) as the scale $\epsilon$ decreases.  As long as no new physical scales are present, such a prediction should be straightforward, and important physical scales can be included by appropriate matching.  This would in a particular way within the context of particle physics, concretely realize an interpretation of machine learning as renormalization group evolution; see, e.g., \InRef{Roberts:2021fes}.

As presented in this paper, however, the compute resources for this smearing analysis may be extreme, even by today's standards.  The number of elements of the distance matrix of $n$ events scales like $n^2$, and so storing the distance matrix for dataset sizes regularly used in modern machine learning studies of, say, $10^7$ events, would require about a petabyte.  Even the SEMD, which can be evaluated extremely fast as far as event metrics go, would still take nearly a billion seconds of wall time to evaluate the distance matrix of $10^7$ events on a modern GPU with a batch size of 10000, unless GPU RAM was significantly increased to allow for larger batches or more parallelization.  As estimated in this paper, however, to really be sensitive to the physics of hadronization within the smearing analysis would require about $10^{10}$ events, pushing all of these requirement estimates to (as of now) almost unfathomable compute resources.  These estimates are of course a quantification of how challenging this problem is, but perhaps the simple idea of smearing over the dataspace can provide insights and progress that renders these estimates irrelevant.

If one allows for more approximation in the evaluation of the event metric, there are likely several ways to reduce compute resources.  For example, as used throughout this paper, all jets were clustered into 100 particles, and the time to evaluate the SPECTER algorithm scales quadratically with the number of particles in the jets.  So, at the cost of loss of resolution of the internal dynamics of the jets, compute times could be decreased by clustering into significantly fewer particles.  Clustering into fewer particles does mean that jet substructure at smaller scales is lost, and this may have a delicate interplay with the minimal or desired smearing resolution $\epsilon$ that one wishes to probe the dataspace manifold.

For some dedicated tasks, there may be ways to focus or hone in on the physics at small distance scales in a compute-resource friendly way.  One possible approach within simulation may be to generate a single event and let the parton shower proceed until some cutoff energy scale, say, $k_{\perp,\text{cut}}$.  Then, from that one event, continue the parton shower and subsequent hadronization many, many times, each with a different random number seed so that radiation at scales lower than $k_{\perp,\text{cut}}$ would be different.  Further, the complete events generated through this procedure would all lie within a metric distance of about $\epsilon \sim k_{\perp,\text{cut}}$ of one another, so the dataspace manifold in this region could be sampled with much higher density than the procedures discussed in this paper.  Such a hyper-focused generation of events in a small neighborhood of the dataspace manifold could be useful for establishing the dimensionality of the space, its local scalar curvature, or other fundamental geometric quantities of interest.  We look forward to application of this smearing approach to many more problems in this growing field.

\section*{Acknowledgments}

I thank Rikab Gambhir for detailed comments, collaboration on related work, and many discussions about event metrics, Yoni Kahn for detailed comments and discussions about scaling laws, and Jesse Thaler for comments.

\appendix

\section{Example Calculations of the Smeared Distributions}

As an example of this procedure, we can straightforwardly calculate the smeared probability distribution of high-energy quark or gluon collinear fragmentation.  For simplicity, we will just work through order-$\alpha_s$, in which the probability distribution on phase space $\Pi$ is
\begin{align}
p(\Pi) &= \delta(s)\delta(z)+\frac{\alpha_s}{2\pi}\frac{1}{s}\, P(z)\,\Theta\left(z(1-z)E^2R^2-s\right)\\
&\hspace{2cm}-\delta(s)\delta(z)\,\frac{\alpha_s}{2\pi} \int\frac{ds'}{s'}\,dz'\, P(z')\,\Theta\left(z'(1-z')E^2R^2-s'\right)+\cdots\nonumber
\end{align}
Here, $P(z)$ is the collinear splitting function \cite{Dokshitzer:1977sg,Gribov:1972ri,Gribov:1972rt,Lipatov:1974qm,Altarelli:1977zs} in energy fraction $z$, and $s$ is the invariant mass of the jet.  The explicit $\Theta$ function constrains the emissions to lie within the jet of radius $R$.  This is a distribution, with $\delta$-functions and other not-functions in its expression.  This introduces problems when we try to calculate ratios of probability distributions to determine the theoretical likelihood.  However, smearing this distribution transmogrifies it into a proper function, of which ratios can be taken and interpreted simply.

On the phase space $d\Pi = ds\, dz$, the $p=2$ SEMD metric takes the form
\begin{align}
d^2(\Pi,\Pi') = 2s+2s'-4\min[z(1-z),z'(1-z')]\sqrt{\frac{s}{z(1-z)}\frac{s'}{z'(1-z')}}\,,
\end{align}
between jets with coordinates $(s,z)$ and $(s',z')$.  With this metric, we can then smear the fragmentation distribution straightforwardly, where
\begin{align}
&\int d\Pi'\, p(\Pi')\,\Theta\left(\epsilon^2-d^2(\Pi,\Pi')\right)=\Theta(\epsilon^2-2s)\\
&\hspace{1cm}+\frac{\alpha_s}{2\pi}\int \frac{ds'}{s'}\, dz'\,P(z')\,\Theta\left(z'(1-z')E^2R^2-s'\right)\nonumber\\
&\hspace{2cm}\times\left[\Theta\left(
\epsilon^2-2s-2s'+4\min[z(1-z),z'(1-z')]\sqrt{\frac{s}{z(1-z)}\frac{s'}{z'(1-z')}}
\right)-\Theta(\epsilon^2-2s)\right]\nonumber\\
&\hspace{1cm}+\cdots\nonumber\,.
\end{align}
While the integrands of the explicit integrals are not necessarily pretty, they are finite and can be evaluated numerically very easily.

\bibliography{refs}

\end{document}